\documentclass[aps,twocolumn,showpacs]{revtex4}
\usepackage{bm}
\usepackage{graphicx}
\usepackage{epsfig}
\begin{document}

\title{Scattering induced spin orientation and spin currents in
gyrotropic structures}
\author{S.\,A.\,Tarasenko}
\affiliation{A.F.~Ioffe Physico-Technical Institute of the Russian
Academy of Sciences, 194021 St.~Petersburg, Russia}
\begin{abstract}
It is shown that additional contributions both to current-induced
spin orientation and to the spin Hall effect arise in quantum
wells due to gyrotropy of the structures. Microscopically, they
are related to basic properties of gyrotropic systems, namely,
linear in the wave vector terms in the matrix element of electron
scattering and in the energy spectrum. Calculation shows that in
high-mobility structures the contribution to the spin Hall current
considered here can exceed the term originated from the Mott skew
scattering.
\end{abstract}

\pacs{72.25.Pn, 72.25.Rb, 73.63.Hs}

\maketitle

\section{Introduction}

Spin-orbit coupling in low-dimensional semiconductor structures is
attracting a great deal of attention since it underlies effects of
manipulating spins of charge carriers by electrical means. Spin
orientation of free carriers by electric
current~\cite{Vorob'ev,Kato,Silov,Ganichev} and the spin Hall
effect, where a charge current drives a transverse spin
current~\cite{DP,Awschalom,Wunderlich}, are among the widely
discussed phenomena in the thriving field of semiconductor
spintronics. Microscopically, both effects are caused by
spin-orbit interaction and can be related either to spin-dependent
scattering (extrinsic contributions) or solely to spin splitting
of the band structure (intrinsic terms). Comparative roles of the
extrinsic and intrinsic mechanisms in spin transport of charge
carriers in bulk and low-dimensional semiconductors are, at
present, the subject of experimental and theoretical discussion
(see Refs.~\cite{Rashba,Sinova} for review).

So far, scattering-induced spin effects in transport of conduction
electrons have been attributed mainly to the Mott skew
scattering~\cite{Rashba,Sinova,Korenev,Side-jump}. The Mott term
in the matrix element of scattering can be written as $\lambda\,
\bm{\sigma}\cdot [\bm{k}\times\bm{k}']$, where $\lambda$ is a
parameter, $\bm{\sigma}$ is the vector of the Pauli matrices,
$\bm{k}$ and $\bm{k}'$ are the initial and scattered wave vectors,
respectively. This term is quadratic in the wave vector and
present in any, even centrosymmetric, structure. In semiconductor
quantum wells (QWs) the Mott term is not dominant contribution to
the spin-dependent part of electron scattering since gyrotropic
symmetry of QW structures allows for linear in the wave vector
coupling of spin states. An example of such a $\bm{k}$-linear
spin-orbit coupling is the Rashba or the Dresselhaus spin-orbit
splitting of the electron subbands induced by Structure and Bulk
Inversion Asymmetry (SIA and BIA, respectively). Similarly, a
spin-dependent term linear in the wave vector appears in the
amplitude of electron scattering by static defects or phonons.
Taking into consideration this contribution and neglecting the
Mott term, the matrix element of electron scattering can be
presented as (see Ref.~\cite{Tarasenko} and references therein)
\begin{equation}\label{V_as}
V_{\bm{k}'\bm{k}} = V_0 + \sum_{\alpha\beta} V_{\alpha\beta} \,
\sigma_{\alpha} (k_{\beta} + k_{\beta}') \:,
\end{equation}
where $V_0$ is the matrix element of conventional spin-conserving
scattering and $V_{\alpha\beta}$ are parameters determined by
space distribution and structure details of the scatterers. In the
case of elastic scattering from short-range static defects, that
is assumed below, the parameters $V_{\alpha\beta}$ and $V_0$ are
independent of the wave vectors $\bm{k}$ and $\bm{k}'$. Linear in
the wave vector term seems to be the dominant spin-dependent
contribution to the matrix element of electron scattering in QWs.
Unlike the Mott contribution, it can be obtained in first order of
the $\mathbf{k} \cdot \mathbf{p}$ perturbation
theory~\cite{Tarasenko}.

In this paper we analyze spin transport of two-dimensional (2D)
electrons in the presence of $\bm{k}$-linear terms in the
scattering amplitude. We show that these terms together with the
spin-orbit spectrum splitting give rise to additional
contributions to both the current-induced spin orientation of free
carriers and the spin Hall effect. The contribution to the spin
orientation of 2D electrons is comparable in magnitude to that
related to current-induced carrier redistribution between the
spin-split subbands~\cite{Aronov}. As regards the spin Hall
current, the proposed contribution can exceed the term originated
from the Mott skew scattering in QW structures with high mobility.

\section{Microscopic model}

Microscopically, mechanism of spin Hall current generation and
spin orientation of 2D electrons by electric current due to
$\bm{k}$-linear terms in the scattering amplitude is as follows.
Application of an electric field $\bm{E}$ in the QW plane  results
in a directed flow of the carriers. Due to spin-dependent
asymmetry of scattering, electrons driven by the electric field
are scattered in preferred directions depending on their spin
states. It leads to generation of a spin-dependent electron
distribution, where particles with a certain wave vector $\bm{k}$
carry a certain spin orientation. The explicit form of the
distribution depends on the origin of the spin-dependent
scattering. Since $\bm{k}$-linear terms in the scattering
amplitude are caused by inversion asymmetry of QWs, one can
distinguish, similarly to the spectrum splitting, the SIA and BIA
contributions to the scattering amplitude. The corresponding
distributions of the spin density in $\bm{k}$ space are shown in
Fig.~1 for electrons confined in (001)-grown QW and subjected to
the in-plane electric field along $[1\bar{1}0]$ and $[110]$ axes.
\begin{figure}[b]
\leavevmode \epsfxsize=0.95\linewidth
\centering{\epsfbox{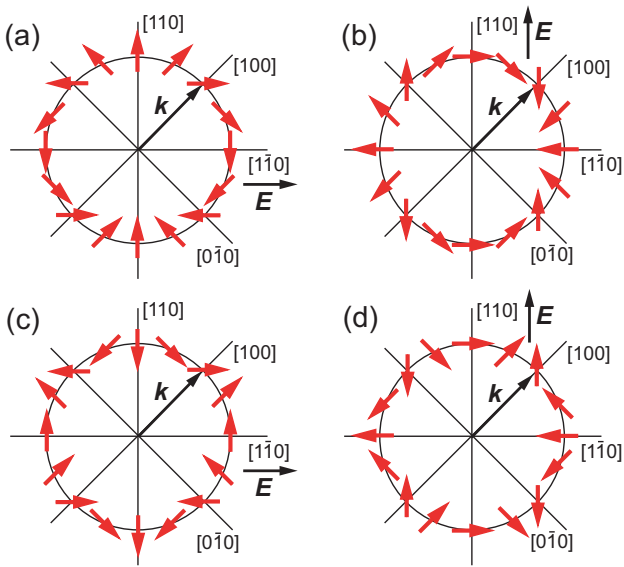}} \caption{Distribution of electron
spin caused by $\bm{k}$-linear terms in the scattering amplitude
for various directions of the electric field. Figs.~(a),(b) and
(c),(d) correspond to the SIA and BIA $\bm{k}$-linear terms in the
scattering amplitude in (001)-grown QWs, respectively.}
\end{figure}
Detailed calculation of the presented distributions is given in
the next section. Here we only note that the angular dependencies
of the spin distributions are described by quadratic harmonics.
Average spin polarization remains zero and, in contrast to the
Mott scattering, electrons with the opposite wave vectors $\bm{k}$
and $-\bm{k}$ carry the same spin. Therefore, the spin-dependent
scattering of the form~(\ref{V_as}) does not lead itself to a
current-induced spin orientation nor to the spin Hall effect.

A net spin orientation of the electron gas and a spin current
appear as a result of the subsequent spin dynamics of carriers.
The spin dynamics of the conduction electrons is known to be
governed by spin-orbit coupling that may be considered as an
effective magnetic field acting on electron spins. In this
effective magnetic field spins of the carriers, initially directed
according to the spin-dependent scattering processes, precess.
Such a spin dynamics is illustrated in Fig.~2.
\begin{figure}[t]
\leavevmode \epsfxsize=0.8\linewidth
\centering{\epsfbox{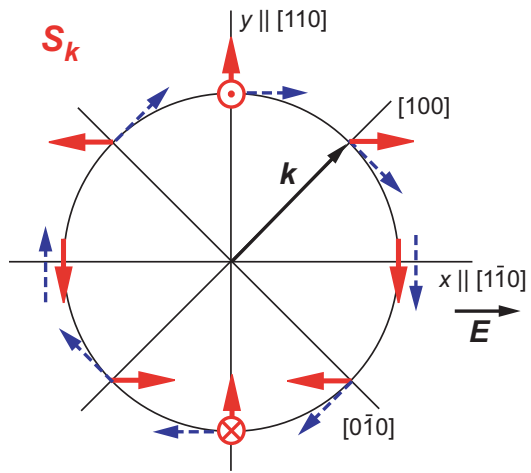}} \caption{Microscopic mechanism of
the current-induced spin orientation and the spin Hall effect.
Spin-dependent asymmetry of electron scattering followed by spin
precession in the effective magnetic field leads to (i) generation
of a spin current and (ii) net spin orientation of carriers.}
\end{figure}
To be specific we consider that the electric field $\bm{E}$ is
directed along the $[1\bar{1}0]$ axis and both the effective
magnetic field and the $\bm{k}$-linear terms in the scattering
amplitude are induced by structure inversion asymmetry. The
effective magnetic field caused by spin-orbit coupling in QWs is
known to be an odd function of the wave vector $\bm{k}$. In the
the case of SIA it is directed perpendicular to $\bm{k}$ as shown
in Fig.~2 by dashed arrows. Therefore, electrons moving along or
opposite to the $x\|[1\bar{1}0]$ axis carry spins oriented
parallel (or antiparallel) to the effective magnetic field, while
the particles moving along or opposite to the $y\|[110]$ axis
carry spins oriented perpendicular to the effective field. As a
result of the spin precession, the spin component $S_z>0$
($S_z<0$) appears for the carriers with the positive (negative)
wave vector $k_y$. This state corresponds to the spin Hall effect
where an electric field drives a transverse spin current, i.e.
oppositely directed flows of carriers with the opposite spins.
Moreover, the spin precession in the effective magnetic field
leads to appearance of an average spin orientation of carriers in
the QW plane, opposite to the $y$ axis. Indeed, electrons with the
spins directed opposite to the $y$ axis are affected by collinear
magnetic field and retain the spin orientation, while the carriers
with the spins directed along the $y$ axis partially lose the
polarization due to the spin precession. The rate of the spin
generation is determined by the average angle of spin rotation in
the effective magnetic field, similarly to the Hanle effect.

\section{Theory}

Theory of the scattering-induced spin orientation and the spin
Hall effect is developed here by using the spin-density-matrix
technique. Dynamics of the density matrix $\rho_{\bm{k}}$ of
electrons subjected to an in-plane electric field $\bm{E}$ is
given by the kinetic equation
\begin{equation}\label{rho}
\frac{\partial \rho_{\bm{k}}}{\partial t} + e \bm{E}
\frac{\partial \rho_{\bm{k}}}{\partial \, \hbar\bm{k}} +
\frac{i}{\hbar}[H_{so},\rho_{\bm{k}}] = \mathrm{St} \rho_{\bm{k}}
\:.
\end{equation}
Here $e$ is the electron charge, $H_{so}$ is the Hamiltonian of
spin-orbit coupling that describes spin precession in an effective
magnetic field
\begin{equation}\label{Hso}
H_{so} = \sum_{\alpha\beta} \gamma_{\alpha\beta} \,\sigma_{\alpha}
k_{\beta} = \frac{\hbar}{2} \, \bm{\Omega}_{\bm{k}} \cdot
\bm{\sigma} \:,
\end{equation}
$\gamma_{\alpha\beta}$ are the structure parameters,
$\bm{\Omega}_{\bm{k}}$ is the Larmor frequency corresponding to
the effective field, and $\mathrm{St} \rho_{\bm{k}}$ is the
collision integral. In the present letter we restrict ourselves to
the case of low temperatures when transport properties are
determined by the Fermi surface and energy mixing of carriers is
suppressed. For elastic scattering from static defects uniformly
distributed in the QW plane the collision integral has the form
\begin{eqnarray}\label{St}
\mathrm{St} \rho_{\bm{k}} = \frac{\pi}{\hbar} N_d \sum_{\bm{k}'}
(2 V_{\bm{k}\bm{k}'} \rho_{\bm{k}'} V_{\bm{k}'\bm{k}} -
V_{\bm{k}\bm{k}'} V_{\bm{k}'\bm{k}} \rho_{\bm{k}} \\
- \rho_{\bm{k}} V_{\bm{k}\bm{k}'} V_{\bm{k}'\bm{k}}) \,
\delta(\varepsilon_{\bm{k}} - \varepsilon_{\bm{k}'}) \:,\nonumber
\end{eqnarray}
where $N_d$ is the sheet density of defects. Note, that spin-orbit
splitting of the energy spectrum is neglected in the collision
integral~(\ref{St}), since the splitting is much smaller than the
electron kinetic energy. Corrections to the energy due to the
spin-orbit splitting as well as similar corrections to the
electron velocity, which can be crucial for intrinsic spin
effects, are unimportant here.

The density matrix can be presented as follows
\begin{equation}
\rho_{\bm{k}} = (f_0 + \delta f_{\bm{k}}) I + \bm{S}_{\bm{k}}
\cdot \bm{\sigma} \:,
\end{equation}
where $f_0$ is the function of the equilibrium carrier
distribution, $I$ is the $2\times2$ unit matrix, $\delta
f_{\bm{k}}$ and $\bm{S}_{\bm{k}}$ are the electric field-induced
corrections to the diagonal and spin components of the density
matrix. We assume that the electric field oscillates at the
frequency $\omega$, $ \bm{E} \propto \exp{(-\mathrm{i}\omega t
)}$. Then, in linear in the electric field regime, the terms
$\delta f_{\bm{k}}$ and $\bm{S}_{\bm{k}}$ have the same time
dependence. To first order in the parameters $V_{\alpha\beta}$ the
charge transport is independent of the spin part of scattering
amplitude and the diagonal correction to the equilibrium density
matrix has the form
\begin{equation}
\delta f_{\bm{k}} = -\frac{ e \tau\, \bm{E} \cdot
\bm{v}}{1-\mathrm{i}\omega\tau} \frac{d f_0}{d \varepsilon} \:,
\end{equation}
where $\bm{v}=\hbar \bm{k}/m^*$ is the velocity, $m^*$ is the
electron effective mass, $\tau=\hbar^3/(V_0^2 N_d\,m^*)$ is the
momentum isotropization time, and $\varepsilon$ is the electron
kinetic energy. Note, that $\delta f_{\bm{k}}$ is the correction
that describes the conventional (Drude) ac conductivity of the
electron gas.

Equation for the spin component of the density matrix
$\bm{S}_{\bm{k}}$ can be derived from Eq.~(\ref{rho}). To first
order in spin-dependent part of the matrix element of scattering
$V_s$ it takes the form
\begin{equation}\label{S_k}
-\mathrm{i}\omega \bm{S}_{\bm{k}} + [\bm{S}_{\bm{k}} \times
\bm{\Omega}_{\bm{k}}] = - \frac{\bm{S}_{\bm{k}} -
\bar{\bm{S}}_{\bm{k}}}{\tau} + \bm{g}_{\bm{k}} \:,
\end{equation}
where $\bar{\bm{S}}_{\bm{k}}$ is $\bm{S}_{\bm{k}}$ averaged over
directions of the wave vector $\bm{k}$, $\bm{g}_{\bm{k}}$ is the
spin generation rate into the state with the wave vector $\bm{k}$,
\begin{equation}\label{g_k}
\bm{g}_{\bm{k}} = \frac{2\pi}{\hbar} N_{d} \sum_{\bm{k}}
\mathrm{Tr}[\bm{\sigma} V_0 V_s] (\delta f_{\bm{k}'} - \delta
f_{\bm{k}}) \,\delta(\varepsilon_{\bm{k}}-\varepsilon_{\bm{k}'})
\:.
\end{equation}
The left-hand side of Eq.~(\ref{S_k}) describes spin dynamics of
the carriers during their free motion between consecutive
collisions with structure defects, while the right-hand side of
the equation stands for the scattering-induced spin redistribution
among electron states with different wave vectors. Particularly,
the first term on the right-hand side of Eq.~(\ref{S_k}) describes
isotropization of the spin density and the second term corresponds
to the spin density generation due to carrier drift in the
electric field in the presence of spin-dependent scattering. We
note, that spin-flip processes are neglected in Eq.~(\ref{g_k})
since they appear only in second order in $V_s$. For linear in the
wave vector terms in the matrix element of scattering given by
Eq.~(\ref{V_as}), components of the spin generation rate have the
form
\begin{equation}\label{g_k2}
g_{\bm{k},\alpha} = \frac{2e \hbar /m^*}{1-\mathrm{i}\omega\tau}
\frac{d f_0}{d \varepsilon} \sum_{\beta \mu}
\frac{V_{\alpha\beta}}{V_0} \left( k_{\beta}k_{\mu} -
\frac{k^2}{2} \delta_{\beta\mu} \right) E_{\mu} \:.
\end{equation}
Dependence of the spin generation $\bm{g}_{\bm{k}}$ on the wave
vector is determined by both the direction of the applied electric
field $\bm{E}$ and the explicit form of the coefficients
$V_{\alpha\beta}$. The latter is governed by the QW symmetry and
can be varied. Symmetry analysis shows that in (001)-grown QWs
there are two nonzero components of the pseudotensor
$V_{\alpha\beta}$, namely $V_{xy}$ and $V_{yx}$, which can be
expressed via the SIA and BIA contributions as follows:
$V_{xy}=V_{BIA} + V_{SIA}$, $V_{yx}=V_{BIA} - V_{SIA}$. Figure~1
demonstrates scattering-induced distributions of the spin
polarization $\bm{g}_{\bm{k}}$ in $\bm{k}$ space for the electric
field directed along the $x$ and $y$ axes. Figures~(a),(b) and
(c),(d) correspond to the cases where the $\bm{k}$-linear terms in
the matrix element of scattering are caused by SIA
($V_{xy}=-V_{yx}$) and BIA ($V_{xy}=V_{yx}$), respectively.

As mention above, the net spin orientation of carriers and the
spin Hall current appear as a result of the spin-dependent
scattering and the subsequent precession of electron spins in the
effective magnetic field. Following Eq.~(\ref{S_k}) and taking
into account that the frequency $\bm{\Omega}_{\bm{k}}$ is an odd
function of the wave vector, while $\bm{g}_{\bm{k}}$ is an even
function of $\bm{k}$ and $\bar{\bm{g}}_{\bm{k}}=0$, one can derive
the equation for the spin component $\bar{\bm{S}}_{\bm{k}}$
\begin{equation}\label{S_aver}
\left\langle \tau
\frac{\bm{\Omega}_{\bm{k}}\times[\bm{\Omega}_{\bm{k}}\times
(\bar{\bm{S}}_{\bm{k}} + \tau\bm{g}_{\bm{k}})]}
{[(1-\mathrm{i}\omega\tau)^2 +(\Omega_{\bm{k}}\tau)^2]}
\right\rangle + \mathrm{i}\omega \bar{\bm{S}}_{\bm{k}} =0 \:,
\end{equation}
where the angle brackets mean averaging over directions of the
wave vector. Solutions of Eqs.~(\ref{S_k})~and~(\ref{S_aver})
allow one to find the spin components of the density matrix and
calculate the spin polarization of the electron gas as well as the
spin Hall current.

\section{Results and discussion}

Calculation shows that in the case $\Omega_{\bm{k}}\tau \ll 1$ the
average electron spin, defined as $\sum_{\bm{k}} \bm{S}_{\bm{k}} /
N_e$, with $N_e$ being the carrier concentration, has the form
\begin{eqnarray}\label{spinorient}
s_x(\omega) = -\frac12
\left[\frac{V_{yx}}{V_0}\frac{\gamma_{xy}}{\gamma_{yx}} +
\frac{V_{xy}}{V_0} \right] \frac{K_{y}(\omega)}{1-\mathrm{i}\omega
T_x(1-\mathrm{i}\omega\tau)^2} \:,\\
s_y(\omega) = -\frac12
\left[\frac{V_{xy}}{V_0}\frac{\gamma_{yx}}{\gamma_{xy}} +
\frac{V_{yx}}{V_0} \right] \frac{K_{x}(\omega)}{1-\mathrm{i}\omega
T_y(1-\mathrm{i}\omega\tau)^2} \:.\nonumber
\end{eqnarray}
Here $V_{xy}$, $V_{yx}$, $\gamma_{xy}$ and $\gamma_{yx}$ are
nonzero components of the pseudotensors $V_{\alpha\beta}$ and
$\gamma_{\alpha\beta}$ in (001)-grown structures, $\bm{K}(\omega)$
is the average electron wave vector corresponding to the drift
velocity of carriers in the electric field
\begin{equation}
\bm{K}(\omega) = \frac{e \tau
/\hbar}{1-\mathrm{i}\omega\tau}\bm{E} \:,
\end{equation}
and $T_{\alpha}$ ($\alpha=x,y$) are the D'yakonov-Perel'
relaxation times of the spin components
\begin{equation}\label{Ts}
1/T_{\alpha} = \tau [\langle\bm{\Omega}_{\bm{k}}^2\rangle -
\langle\Omega_{\bm{k},\alpha}^2\rangle] \:.
\end{equation}
Magnitude of the spin orientation achieved by dc electric current
$s(0)$ depends on the ratio $\gamma_{xy}/\gamma_{yx}$ rather than
on absolute value of the spin-orbit splitting
$\hbar\Omega_{\bm{k}}$. This can be understood considering the
average spin as a balance between processes of spin generation and
spin relaxation. $s(0)$ is given by $T\dot{s}$, where the spin
relaxation time $T\propto1/\Omega_{\bm{k}}^2$ and the spin
generation rate $\dot{s}\propto\Omega_{\bm{k}}^2$.

The spin current is characterized by a pseudotensor $\hat{\bm{J}}$
with the components $J_{\beta}^{\alpha}$ describing the flow in
the $\beta$ direction of spins oriented along the $\alpha$ axis.
In terms of the kinetic theory such a component of the spin
current is contributed by a non-equilibrium correction $\propto
\sigma_{\alpha} k_{\beta}$ to the electron spin density matrix and
given by
\begin{equation}
J_{\beta}^{\alpha} = \sum_{\bm{k}} \mathrm{Tr} \left[
\frac{\sigma_{\alpha}}{2} v_{\beta} \rho_{\bm{k}} \right] =
\sum_{\bm{k}} S_{\bm{k},\alpha} v_{\beta} \:.
\end{equation}
Calculation shows that the components of the scattering-induced
spin Hall current in (001)-grown QWs have the form
\begin{eqnarray}\label{spincurrent}
J_{x}^{z}(\omega) &=& \mathrm{i} \frac{\tau k_F^2}{2\hbar}
\left[\frac{V_{yx}}{V_0}\gamma_{xy} +
\frac{V_{xy}}{V_0}\gamma_{yx} \right] \frac{\omega
T_x\,j_y(\omega)/e}
{1-\mathrm{i}\omega T_x(1-\mathrm{i}\omega\tau)^2} \:,\\
J_{x}^{z}(\omega) &=& -\mathrm{i} \frac{\tau k_F^2}{2\hbar}
\left[\frac{V_{yx}}{V_0}\gamma_{xy} +
\frac{V_{xy}}{V_0}\gamma_{yx} \right] \frac{\omega
T_y\,j_x(\omega)/e} {1-\mathrm{i}\omega
T_y(1-\mathrm{i}\omega\tau)^2} \:,\nonumber
\end{eqnarray}
where $k_F$ is the Fermi wave vector and $\bm{j}(\omega)=e N_e
\hbar \bm{K}(\omega)/m^*$ is the charge current.

Equations~(\ref{spinorient})~and~(\ref{spincurrent}) describe the
current-induced spin orientation and the spin Hall current for the
case $\Omega_{\bm{k}}\tau \ll 1$. In high-mobility 2D structures
this inequality may be violated. At arbitrary
$\Omega_{\bm{k}}\tau$ solutions of
Eqs.~(\ref{S_k})~and~(\ref{S_aver}) have complicated form but
become simpler if the absolute value of the Larmor frequency
$\Omega_{\bm{k}}$ is independent of the wave vector direction.
This is fulfilled when the spin-orbit splitting is caused only by
BIA or by SIA, $\gamma_{xy}/\gamma_{yx}=\pm 1$, respectively. In
this particular case the spin orientation and the spin Hall
current are given by
Eqs.~(\ref{spinorient})~and~(\ref{spincurrent}), where the
denominators $1-\mathrm{i}\omega
T_{\alpha}(1-\mathrm{i}\omega\tau)^2$ are replaced by
$1-\mathrm{i}\omega T[(1-\mathrm{i}\omega\tau)^2+2\tau/T]$, with
the time $T$ ($T_x=T_y$) being determined by Eq.~(\ref{Ts}).

Frequency dependencies of the current-induced spin polarization
$|s_y(\omega)|$ and the spin Hall current
$|eJ_y^z(\omega)/j_x(0)|$ are plotted in Fig.~3 for different
parameters $\Omega_{k_F}\tau$. It is assumed that the electric
field is directed along the $x$ axis and the $\bm{k}$-linear terms
both in the scattering amplitude and in the spectrum splitting are
related to structure inversion asymmetry. Magnitude of the spin
orientation reaches maximum in the dc limit and decreases with
increasing the field frequency $\omega$, see Fig.~3(a). In systems
where $\Omega_k\tau \ll 1$, the spin relaxation time is much
longer than the momentum relaxation time, $T\gg\tau$, and the
frequency dependence of the spin orientation $s_{\alpha}(\omega)$
is given by $s_{\alpha}(0)/(1-\mathrm{i}\omega T_{\alpha})$. Such
a behavior is natural because the spin orientation represents a
carrier redistribution between the spin states and the spin
dynamics is governed by the spin relaxation time.
\begin{figure}[t]
\leavevmode \epsfxsize=0.9\linewidth
\centering{\epsfbox{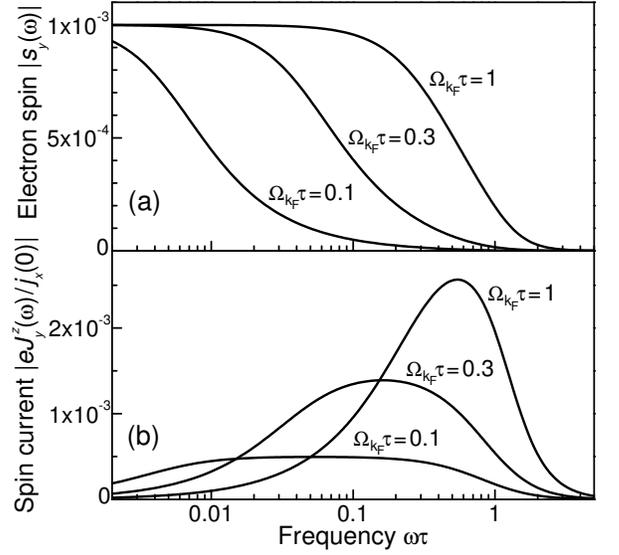}} \caption{Frequency dependencies of
(a) the spin orientation $|s_y(\omega)|$ and (b) the spin current
$|eJ_y^z(\omega)/j_x(0)|$ for different $\Omega_{k_F}\tau$ and
$V_{xy}/V_0=10^{-8}$~cm, $k_F=10^6$~cm$^{-1}$,
$K(0)=10^5$~cm$^{-1}$.}
\end{figure}

In contrast to the current-induced spin orientation, the spin Hall
current reveals nonmonotonic frequency dependence. It vanishes at
zero frequency, increases linearly with the frequency at low
$\omega$, reaches maximum and decreases when $\omega$ exceeds
$1/\tau$. The cancellation of the spin Hall effect in the dc limit
as well as the linear growth at small $\omega$ is a direct
consequence of the particular relation between spin dynamics and
spin fluxes in systems with spin-orbit splitting of the band
structure linear in the wave vector~\cite{Dimitrova}. The relation
between components of the spin and the spin current can be easily
obtained from Eq.~(\ref{S_k}). Summing Eq.~(\ref{S_k}) over
$\bm{k}$ and taking into account that $\Omega_{\bm{k}}$ is a
linear function of the wave vector, in particular,
$\Omega_{\bm{k},x}\propto k_y$ and $\Omega_{\bm{k},y}\propto k_x$
in (001)-grown structures, one obtains the relation
$J_{\alpha}^{z}(\omega)\propto\omega s_{\alpha}(\omega)$. At small
frequencies $\omega$ the spin orientation is a finite value,
implying the linear growth of the spin Hall current with the
frequency. In the range $1/T \ll \omega \ll 1/\tau$ the spin
orientation drops as $\omega^{-1}$ and the frequency dependence of
the spin Hall current has a plateau. After that, in the
high-frequency limit, $\omega \gg 1/\tau$, the spin orientation
and the spin Hall current decrease as $\omega^{-4}$ and
$\omega^{-3}$, respectively. We note that in the high-frequency
range the absorption of ac electric field by free carriers can be
significant and result in pure spin
photocurrents~\cite{Spinflux,Belkov} and spin orientation of the
carriers~\cite{Tarasenko}.

Finally, we present estimations for the considered effects.
Following Eq.~(\ref{spinorient}) the current-induced spin
orientation at zero frequency can be estimated as $10^{-3}$ for
$V_{xy}/V_0\sim10^{-8}$~cm, appropriate to GaAs-based structures,
and the drift wave vector $K(0)\sim10^5$~cm$^{-1}$. This value
corresponds to the spin orientation caused by other mechanism,
namely, current-induced electron redistribution between the
spin-split subbands. For this particular mechanism the spin
orientation can be estimated as $\gamma_{xy}K(0)/E_F$ that also
gives $10^{-3}$ for $\gamma_{xy}\sim10^{-7}$~meV$\cdot$cm and the
Fermi energy $E_F\sim10$~meV~\cite{Aronov}. Further increase of
the Fermi energy (or temperature for the Boltzmann statistics)
leads to predominance of the scattering-related mechanism. As
regards the spin Hall effect, the ratio of the spin
current~(\ref{spincurrent}) to the contribution caused by the Mott
skew scattering can be estimated as $V_{xy}/(\lambda k_F)\,
\Omega_{k_F}\tau$. In high mobility structures, where
$\Omega_{k_F}\tau \sim 1$, the spin Hall current caused by ${\bm
k}$-linear terms in the scattering amplitude exceeds by an order
of magnitude the contribution related to the skew scattering.

\paragraph*{Acknowledgement.} This work was supported by the RFBR,
President Grant for Young Scientists, Russian Science Support
Foundation, and Foundation ``Dynasty'' - ICFPM.


\begin{thebibliography}{99}
\bibitem{Side-jump} The side-jump contribution to the spin Hall current appears in
next order in the parameter $1/(k_Fl)$, where $k_F$ is the Fermi
wave vector and $l$ in the mean free path~\cite{Rashba}.
\end{thebibliography}
\end{document}